\documentclass[aps,prl,showpacs,floatfix,twocolumn]{revtex4-1}
\usepackage{times}
\usepackage{graphicx,graphics,color}
\usepackage{amsmath}
\usepackage{amssymb}
\usepackage{color}

\begin{document}

\title{Quasiparticle Scattering Interference  in  (K,Tl)Fe$_{x}$Se$_{\rm 2}$ Superconductors}

\author{Jian-Xin Zhu}
\email[To whom correspondence should be addressed. \\ Electronic address: ]{jxzhu@lanl.gov}
\homepage{http://theory.lanl.gov}
\affiliation{Theoretical Division, Los Alamos National Laboratory,
Los Alamos, New Mexico 87545}


\author{A. R. Bishop}
\affiliation{Theoretical Division, Los Alamos National Laboratory,
Los Alamos, New Mexico 87545}

\begin{abstract}
We model the quasiparticle interference (QPI) pattern  
in the recently discovered (K,Tl)Fe$_{x}$Se$_{\rm 2}$ superconductors. 
We show in the superconducting state that, due to the absence of hole pockets at the Brillouin zone center, the quasiparticle scattering occurs around the momentum transfer $\mathbf{q}=(0,0)$ and $(\pm \pi, \pm \pi)$ between electron pockets located at the zone boundary. More importantly,  although both $d_{x^2-y^2}$-wave 
and $s$-wave pairing symmetry lead to nodeless quasiparticle excitations, 
distinct QPI features are predicted between both types of pairing symmetry.
 In the presence of a nonmagnetic 
impurity scattering, the QPI exhibits strongest scattering with $\mathbf{q}=(\pm \pi, \pm \pi)$ for the $d_{x^2-y^2}$-wave pairing symmetry; while the strongest scattering 
exhibits a ring-like structure centered around both $\mathbf{q}=(0,0)$ and $(\pm \pi, \pm \pi)$ for the isotropic $s$-wave pairing symmetry.  A unique  QPI pattern has also been predicted due to a local pair-potential-type impurity scattering. 
The significant contrast in the QPI pattern between the $d_{x^2-y^2}$-wave and the isotropic $s$-wave pairing symmetry can be used to probe the pairing symmetry  within the Fourier-transform STM technique. 

 \end{abstract}
\maketitle

{\it Introduction.~} The very recent discovery of high-$T_c$ (above 30 K) superconductivity in
AFe$_{x}$Se$_2$ (A= K, Tl, Cs)~\cite{JGuo:2010,AKMaziopa:2011,MFang:2010} has generated new excitement in the condensed matter community.
 Relative to other iron-based superconductors (such as LaOFeAs, BaFe$_2$As$_2$, FeSe etc.), the end members, TlFe$_2$Se$_2$ and KFe$_2$Se$_2$ (called as the ``122'' iron-selenides), are heavily electron doped (0.5 electron/Fe). Band structure calculations~\cite{LZhang:2009,XWYan:2010b,CCao:2010,IShein:2010,INekrasov:2011}
 on these end compounds show only electron pockets, primarily located
  around the $M$ point of the Brillouin zone (BZ) defined for a simple tetragonal structure.
Angle-resolved photoemission spectroscopy (ARPES) measurements
observed these electron-like Fermi surface pockets around the $M$ points, and showed
no hole-like pockets~\cite{TQian:2010,YZhang:2010} but very weak
electron-like pockets~\cite{DMou:2011,XPWang:2011,LZhao:2011} near the zone center $\Gamma$. The strong Fe-deficient compound ($x \leq 1.6$) shows insulating behavior~\cite{MFang:2010,DMWang}. This  is  in contrast to other iron-based parent compounds, which are poor metals,
 raising the interest in the possibility of a Mott insulating state~\cite{XWYan:2010a, CCao:2011,RYu:2011,YZhou:2011} induced by patterned 
 Fe-vacancies~\cite{ZWang:2011,LHaggstrom:1991}.
These observations add to the possibility that the pairing symmetry in the new compounds 
is unconventional. In particular, the absence of $\Gamma$-centered hole pockets would 
invalidate the popular $s_{\pm}$-type of pairing symmetry, which was proposed for earlier 
iron-based superconductors. Recent calculations have predicted that the superconducting state could have $d_{x^2-y^2}$-wave~\cite{FWang:2011,TMaier:2011,TDas:2011,RYu:2011b} 
and $s$-wave symmetry~\cite{YZhou:2011}.
All these scenarios lead to nodeless superconducting gap structure, which is
 in agreement with the ARPES observations  and other experiments.
Because of the particular Fermi surfaces, conventional phase-sensitive measurements
cannot be readily applied to differentiate the pairing states. 

Recently, one of the present authors and co-workers have proposed use of the existence or absence of intra-gap 
resonance states induced by a nonmagnetic impurity to probe the superconducting pairing 
symmetry~\cite{JXZhu:2011}. It has been found that the impurity-induced resonance state can only 
exist for a $d_{x^2-y^2}$-wave pairing state. As mentioned above, since this $d_{x^2-y^2}$-wave 
pairing state does not introduce nodal quasiparticles, due to the unique Fermi surface topology, 
the impurity-induced resonance state 
is located near the superconducting gap edge and requires a strong potential scattering. 
An alternative technique, which can directly identify the sign change of the superconducting order parameter across various regions of Fermi surface, is 
 the quasiparticle interference (QPI) probe. 
This technique has made a great stride in understanding low-energy quasiparticle properties and superconducting gap symmetry in high-$T_c$ cuprates~\cite{JEHoffman:2002,
QHWang:2003,DZhang:2003}. The underlying principle for the QPI is that even a weak impurity scattering will mix two electronic states with two different momenta but on the same shell energy contour in the Brillouin zone, and the resultant momentum transfer (or scattering interference pattern) corresponds to the modulation in the local density of states, which can be measured by the Fourier-transform STM technique~\cite{AVBalatsky:2006}.  The analysis of QPI in the presence of impurity scattering has been theoretically proposed~\cite{FWang:2009,YYZhang:2009,} to probe the pairing symmetry in earlier iron-based superconductors. Results of later QPI  measurements 
~\cite{THanaguri:2010} are 
consistent with the scenario of the order parameter having a sign reversal across the electron  and hole pockets. Recently, the QPI signatures have also been 
discussed either for the whole phase diagram including the metallic spin-density wave order~\cite{AAkbari:2010} or in the presence of magnetic field~\cite{SSykora:2011}.

Motivated by this recent success, here we perform a detailed analysis of QPI in the newly
discovered ``122'' iron-selenide superconductors.  Both the nonmagnetic impurity scattering  and pair potential scattering are considered. The latter type of scattering is more relevant to the STM experiments on samples with an applied magnetic field, in which Abrikosov vortices are generated. We show a pronounced difference in the QPI characteristic of a simple $s$-wave and 
$d_{x^2-y^2}$-wave pairing symmetry. Because the Fermi surface of the new (K,Tl)Fe$_{x}$Se$_{2}$ compounds comprise small pockets of only one type of carrier,
this kind of study will also provide an opportunity to identify the unique role of Fermi surface 
topology in the QPI pattern even for the same pairing symmetry.  

{\em QPI methodology.~} In view of the fact that there is either no hole pocket or faint features of electron pockets in the 
superconducting  ``122'' iron selenides, we consider here a simplified 
single-band model, which enables us to obtain a full understanding of the QPI pattern 
due to the different Fermi surface topology and pairing symmetry.
 Recently, a similar simplified single-band model (Hubbard model)
was also used to understand the magnetism in the iron-deficient compounds~\cite{HChen:2011}. The model Hamiltonian is defined on a two-dimensional (2D) square lattice and consists
of the pristine and impurity scattering parts, $H=H_0+H_{\text{imp}}$. The pristine part 
$H_0$ is,
\begin{equation}
H= \sum_{\mathbf{k},\sigma} 
\xi_{\mathbf{k}} c_{\mathbf{k}\sigma}^{\dagger} c_{\mathbf{k}\sigma}
+\sum_{\mathbf{k}} [\Delta_{\mathbf{k}} c_{\mathbf{k}\uparrow}^{\dagger} 
c_{\mathbf{k}\downarrow}^{\dagger} + H.c.] \;.
\end{equation} 
 Here the operators
 $c_{\mathbf{k}\sigma}^{\dagger}$ ($c_{\mathbf{k}\sigma}$) create (annihilate) an electron 
 with momentum $\mathbf{k}$ and spin $\sigma$. The quantity $\xi_{\mathbf{k}}$ denotes 
 the energy dispersion.
 We consider only the spin-singlet pairing and the superconducting gap function is described by $\Delta_{\mathbf{k}}$.
The impurity scattering part is given by
\begin{equation}
H_{\text{imp}}= U_0 \sum_{\sigma} c_{0\sigma}^{\dagger} c_{0\sigma} + \delta\Delta \sum_{\delta} 
\eta_{\delta} [c_{0\uparrow}^{\dagger}c_{\delta\downarrow}^{\dagger} + c_{\delta\uparrow}^{\dagger} c_{0\downarrow}^{\dagger} + \text{H.c.}]\;,
\end{equation}
where the first term represents the zero-range normal potential scattering from a nonmagnetic 
impurity located at the origin of the lattice (for mathematical convenience) and the scattering strength is denoted as $U_0$, while the 
second term describes the electron scattering due to a perturbation in the pair potential of  amplitude $\delta\Delta$.  For $s$-wave pairing symmetry, $\eta_{\delta}=1$ with 
the variable $\delta$ in the summation taking only the value of zero;  while for $d$-wave pairing symmetry $\eta_{\delta} = 1\; (-1)$ for $\delta=\pm \hat{x} \;(\pm \hat{y})$. 
For our purpose here, we calculate the Green's function in the presence of either a nongmagnetic impurity or pair potential impurity scattering, which is defined as $\hat{G}(ij;\tau)=-\langle T_{\tau}[\Psi_{i}(\tau)
\Psi_{j}^{\dagger}]\rangle$ with $\Psi_{i}^{\dagger}=(c_{i\uparrow}^{\dagger}, c_{i\downarrow})$
as a two-component operator in the Nambu space. For the weak scattering potential, we make a first-order approximation and obtain
 the Green's function 
\begin{eqnarray}
\hat{G}(i,j;i\omega_n) &=& \hat{G}_{0}(i,j;i\omega_n) + U_0 \hat{G}_{0}(i,0;i\omega_n) \hat{\tau}_{3} \hat{G}_{0}(0,j;i\omega_n)  \nonumber \\
&&+\delta\Delta  \sum_{\delta} \eta_{\delta} \hat{G}_{0}(i,0;i\omega_n)\hat{\tau}_{1} \hat{G}_{0}(\delta,j;i\omega_n)\;,
\end{eqnarray}
where $\omega_n=(2n+1)\pi T$ 
is the Matsubara frequency with $n$  an integer and  $T$ the electronic temperature, $\tau_1$ and $\tau_3$ are the components of Pauli matrices 
in the Nambu space, while $\hat{G}_0$ is the Green's function for the pristine system. From now on, we will term the usual nonmagnetic impurity scattering $\tau_3$-scattering and the scattering off a pair potential impurity $\tau_1$-scattering. We emphasize again that the latter type of scattering is more relevant to the STM measurements 
of samples in the presence of vortices. 
The QPI is characterized by the Fourier-transform of the local density of states (LDOS), that is, 
$\rho_{\mathbf{q}}(E)=\sum_{i} \rho_{i}(E) e^{-i\mathbf{q}\cdot \mathbf{r}_i}$. 
Here the LDOS  is given by $\rho_{i}(\omega) = -(2/\pi)\text{Im}[\delta G_{11}(i,i;i\omega_n\rightarrow E+i\gamma)]$,
with $G_{11}$ being the 
site-diagonal normal (electron) component of the matrix Green's function.
Note that we have measured the LDOS in the presence of the impurity scattering by removing the uniform background. A little algebra yields~\cite{JXZhu:2006}:
\begin{eqnarray} 
\rho_{\mathbf{q}}(E) & = & \frac{U_0}{N_L} \sum_{\mathbf{k}} \{[A_{\mathbf{k}}(E)
B_{\mathbf{k}+\mathbf{q}}(E) + B_{\mathbf{k}}(E)A_{\mathbf{k}+\mathbf{q}}(E)]
 \nonumber \\
&& -[J_{\mathbf{k}}(E) K_{\mathbf{k}+\mathbf{q}}(E)+K_{\mathbf{k}}(E)
J_{\mathbf{k}+\mathbf{q}}(E)]\} 
\nonumber \\
&& + \frac{\delta\Delta}{N_L} \sum_{\mathbf{k}} F_{\mathbf{k}} 
\{[A_{\mathbf{k}}(E)  K_{\mathbf{k}+\mathbf{q}}(E)   +  K_{\mathbf{k}}(E) A_{\mathbf{k}+\mathbf{q}}(E) ]
\nonumber \\
&&  +  [B_{\mathbf{k}}(E)  J_{\mathbf{k}+\mathbf{q}}(E) 
+J_{\mathbf{k}}(E) B_{\mathbf{k}+\mathbf{q}}(E) 
]\}\;,
\label{EQ:rhoq}
\end{eqnarray}
where the form factor $F_{\mathbf{k}}=2$ for $s$-wave pairing symmetry, while 
$F_{\mathbf{k}}=2(\cos k_x - \cos k_y)$ for $d$-wave pairing symmetry. The functions $A$, $B$, 
$J$, and $K$ are defined as $A(J) = -(2/\pi)\text{Im} [G_{0,11(12)}(\mathbf{k};E)]$ and 
$B(K)=\text{Re}[G_{0,11(12)}(\mathbf{k};E)]$.
 Equation~(\ref{EQ:rhoq}) shows that the QPI pattern 
is determined by the convolution of the bare Green's functions 
in momentum space. For $\tau_3$-scattering, the convolution occurs between either normal Green's functions or anomalous ones (with a negative sign); 
while for $\tau_1$-scattering, the convolution is between normal and anomalous Green's functions. These bare Green's functions are given by
\begin{subequations}
\begin{eqnarray}
G_{0,11}(\mathbf{k};E) &=& \frac{u_{\mathbf{k}}^{2}}{E-E_{\mathbf{k}}} + \frac{v_{\mathbf{k}}^{2}}{E+E_{\mathbf{k}}} \;, \\
G_{0,12}(\mathbf{k};E) &=& u_{\mathbf{k}} v_{\mathbf{k}} \biggl{[} \frac{1}{E-E_{\mathbf{k}}} - \frac{1}{E+E_{\mathbf{k}}}\biggr{]}\;,
\end{eqnarray}
\end{subequations}
where the quasiparticle energy $E_{\mathbf{k}} = \sqrt{\xi_{\mathbf{k}}^2 +
\Delta_{\mathbf{k}}^2}$, and  $u_{\mathbf{k}}=\sqrt{(1+\xi_{\mathbf{k}}/E_{\mathbf{k}})/2}$ and 
$v_{\mathbf{k}}=\text{sign}(\Delta_{\mathbf{k}})\sqrt{(1- \xi_{\mathbf{k}}/E_{\mathbf{k}})/2}$  
are the electron and hole parts of the Bogoliubov wavefunction amplitude. Therefore, for a given momentum $\mathbf{k}$, the contribution to the Fourier amplitude $\rho_{\mathbf{q}}(E)$ is sensitive to the sign of $v_{\mathbf{k}} v_{\mathbf{k}+\mathbf{q}}$, which holds the key to reveal the uniqueness of a superconducting pairing symmetry in the QPI measurements.

\begin{figure}[t]
\centering\includegraphics[width=1.0\linewidth]{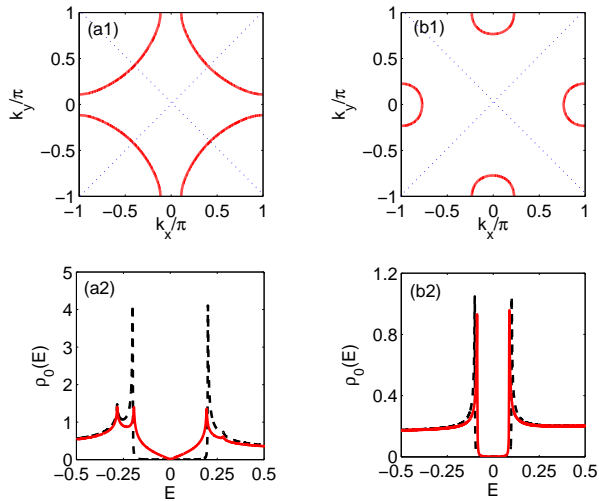}
\caption{(Color online) Fermi surface contour and density of states with $t=1$, $t^{\prime}=-0.3$, and $\mu=-1$ (a) and $t=0$, $t^{\prime}=-1$, and $\mu=-3$ (b), which are relevant to the cuprates and ``122'' iron selenides, respectively. In panels (a2) and (b2), the solid lines are for $d_{x^2-y^2}$-wave pairing symmetry ($\Delta_{\mathbf{k}}=\Delta_0(\cos k_x - \cos k_y)/2)$ with $\Delta_0=0.2$), while the dashed lines are for isotropic $s$-wave pairing symmetry ($\Delta_{\mathbf{k}}=\Delta_0$ with $\Delta_0=0.1$). The dashed lines in panels (a1) and (b1) represent the diagonals across the BZ.
}
\label{FIG:DISP}
\end{figure}

{\em QPI in $d_{x^2-y^2}$-wave and $s$-wave pairing symmetry.~} Before we present the numerical results on the QPI, we point out that, for unconventional pairing symmetry, the  
quasiparticle energy is very sensitive to the Fermi surface topology. To illustrate this point, 
we consider the normal state energy dispersion:
\begin{equation} 
\xi_{\mathbf{k}} = -2t (\cos k_x + \cos k_y)  -4t^{\prime} \cos k_x \cos k_y -\mu\;,
\end{equation} 
where $t$ and $t^{\prime}$ are the nearest-neighbor and next-nearest-neighbor hopping integrals 
and $\mu$ is the chemical potential. If, as relevant to high-$T_c$ cuprates, we take  $t=1$,  $t^{\prime}=-0.3$, and $\mu=-1.0$, the Fermi pockets are centered around the $(\pi,\pi)$ and equivalents
 in the BZ, 
as shown in Fig.~\ref{FIG:DISP}(a1). These pockets are cut by the zone diagonals, which 
makes the quasiparticle excitations gapless in the $d_{x^2-y^2}$-wave pairing symmetry, 
$\Delta_{\mathbf{k}}=(\Delta_0/2)(\cos k_x - \cos k_y)/2$, and a $V$-shaped profile of quasiparticle density 
of states around the Fermi energy (see Fig.~\ref{FIG:DISP}(a2)). It enables the analysis of QPI in cuprates to provide a detailed band structure and $d_{x^2-y^2}$-wave gap structure~\cite{JEHoffman:2002,QHWang:2003,DZhang:2003}.  However, if we take
$t=0$, $t^{\prime}=-1$, and $\mu=-3$, as a simplified modeling of the band structure in 
the ``122'' iron selenides, the Fermi pockets are centered around the $(\pi,0)$ point and equivalents in the BZ (see Fig.~\ref{FIG:DISP}(b1)). Due to the low density of electrons, these Fermi 
pockets are small in cross-section and, as revealed experimentally, are nearly isotropic. 
In this situation, the quasiparticle excitations are fully gapped and a well-shaped profile
of density of states is obtained (see Fig.~\ref{FIG:DISP})(b2)). For the isotropic $s$-wave pairing symmetry, the quasiparticle excitations are always fully gapped,  
irrespective of the detailed Fermi surface topology (see Fig.~\ref{FIG:DISP}(a2)-(b2)). 
This explains why both $d_{x^2-y^2}$-wave 
and isotropic $s$-wave pairing symmetry scenarios are competing candidates
for the photoemission spectroscopy measurements in ``122'' iron selenides. 
To clarify the pairing symmetry, the QPI would be a powerful technique.

We now turn to the QPI by focusing on the special features 
arising from the $d_{x^2-y^2}$-wave and isotropic $s$-wave pairing symmetries.
We fix the band structure parameters with values of $t=0$, $t^{\prime}=-1$, and $\mu=-3$,
with the maximum pair potential amplitude $\Delta_0=0.1$. A lattice size of $N_{L}=2048 \times 2048$ is typically taken. The strength of the $\tau_3$- and $\tau_1$-scattering potentials are taken
as $U_0=0.1$ and $\delta \Delta=0.01$, respectively.  
Because for both $d_{x^2-y^2}$-wave and isotropic 
$s$-wave pairing symmetry with the type of Fermi surface for  ``122'' iron selenides, the quasiparticle excitations are fully gapped, we take the energy $E=\pm 0.1$, which is close to 
the coherent gap edge, for the QPI analysis. The intrinsic broadening parameter is taken as 
$\gamma = 0.02$.

\begin{figure}[t!]
\centering\includegraphics[
width=1.0\linewidth,clip]{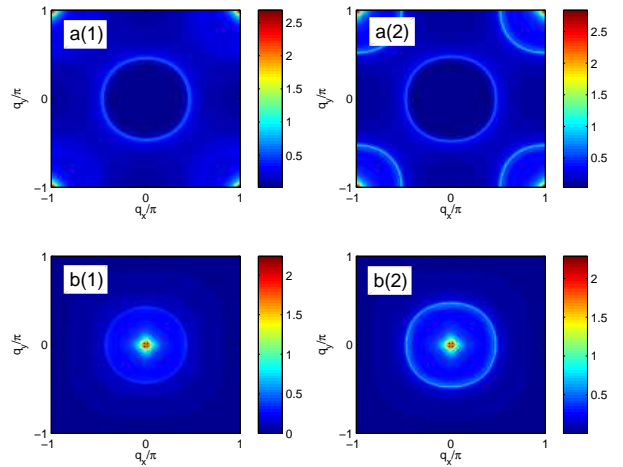}
\caption{(Color online) The quasiparticle scattering map, $\vert \rho_{\mathbf{q}}(E) \vert $, for 
 $\tau_3$ impurity scattering (upper panels) and $\tau_1$  impurity scattering (lower panels) at energy $E=-0.1$ (left column) and $E=0.1$ (right column) with the nodeless $d_{x^2-y^2}$-wave pairing symmetry.  For the $\tau_3$ impurity scattering, the intensity is amplified by a factor of 10, while for the $\tau_1$ impurity scattering, it is amplified by a factor of 25.
}
\label{FIG:QPI_d}
\end{figure}

In Fig.~\ref{FIG:QPI_d}, we show the quasiparticle scattering pattern $\vert \rho_{\mathbf{q}}(E) \vert $ in the presence of either $\tau_3$-scattering or $\tau_1$-scattering for the nodeless $d_{x^2-y^2}$-wave pairing symmetry.   As can be seen  from Fig.~\ref{FIG:QPI_d} (a1)-(a2),  for the $\tau_3$-scattering, the QPI pattern exhibits a ring-like structure around momentum-transfer
 $\mathbf{q}=(0,0)$ and 
$\mathbf{q}=(\pm \pi,\pm \pi)$. However, the minimal intensity occurs at $\mathbf{q}=(0,0)$, while 
the maximum intensity occurs at $\mathbf{q}=(\pm \pi,\pm \pi)$.  On the one hand,  the maximum intensity at $\mathbf{q}=(\pm \pi,\pm \pi)$ indicates strongest scattering between the Fermi pockets located at  $(\pm \pi,0)$ and $(0,\pm \pi)$, which is due to the opposite signs of the superconducting order parameter around these two Fermi pockets. Furthermore, the ring-like structure centered around $\mathbf{q}=(\pm \pi,\pm \pi)$ for $-\vert E \vert$ is much weaker in intensity than that for 
$\vert E\vert$. On the other hand,  the ring-like structure around $\mathbf{q}=(0,0)$ is contributed from the intra-pocket scattering.  In the presence of $\tau_1$-scattering, the quasiparticle scattering changes significantly when compared with the case of $\tau_3$-scattering.
In this case, the bright ring-structure only occurs around $\mathbf{q}=(0,0)$, while almost no 
measurable intensity is obtained around $\mathbf{q}=(\pm \pi,\pm \pi)$. In particular, the maximum 
intensity is located at the point $\mathbf{q}=(0,0)$. The significant difference in the quasiparticle 
scattering pattern between the cases of $\tau_3$-scattering and $\tau_1$-scattering suggests that
the QPI is also very sensitive to the nature of the impurity scattering. Therefore,
in the interpretation of experimental data for superconducting pairing symmetry, 
caution must be taken as to whether the impurity scattering is of $\tau_3$ or $\tau_1$ nature. 

\begin{figure}[t!]
\centering\includegraphics[
width=1.0\linewidth,clip]{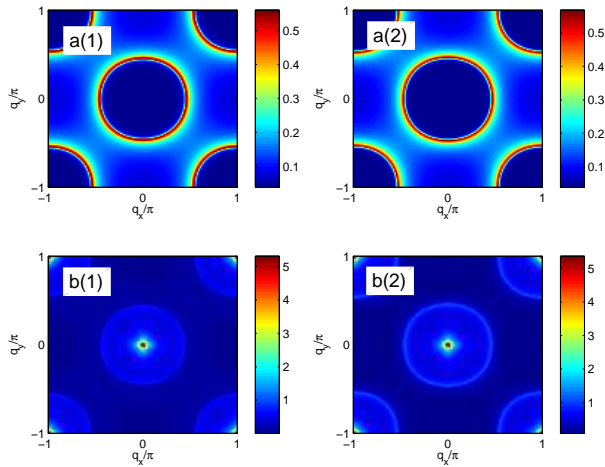}
\caption{(Color online)
The QPI maps for the $\tau_3$ impurity scattering (upper panels) and $\tau_1$  impurity scattering (lower panels) at energy $E=-0.1$ (left column) and $E=0.1$ (right column) with isotropic $s$-wave pairing symmetry.  For  $\tau_3$ impurity scattering, the intensity is amplified by a factor of 10, while for  $\tau_1$ impurity scattering it is amplified by a factor of 50.
}
\label{FIG:QPI_s}
\end{figure}

In Fig.~\ref{FIG:QPI_s}, we show the quasiparticle scattering pattern $\vert \rho_{\mathbf{q}}(E) 
\vert$ in the presence of either $\tau_3$-scattering or $\tau_1$-scattering for isotropic $s$-wave pairing symmetry.  For $\tau_3$-scattering (see Fig.~\ref{FIG:QPI_s}(a1)-(a2)), the ring-like structure appears around both $\mathbf{q}=(0,0)$ and $(\pm \pi,\pm \pi)$. However, in striking contrast to the case of $d_{x^2-y^2}$-wave pairing symmetry (in the same condition of 
$\tau_3$-scattering), the maximum intensity is located directly on the rings, while almost no visible intensity is obtained at both $\mathbf{q}=(0,0)$ and $(\pm \pi,\pm \pi)$  points.  
For $\tau_1$-scattering (see Fig.~\ref{FIG:QPI_s}(b1)-(b2)), the ring-like structure remains centered around the $\mathbf{q}=(0,0)$ and $\mathbf{q}=(\pm \pi,\pm \pi)$ points. More noticeably, the maximum intensity is now located at both $\mathbf{q}=(0,0)$ and $(\pm \pi,\pm \pi)$ points, 
which is also significantly different than the case of $d_{x^2-y^2}$-wave pairing symmetry, 
where the maximum intensity occurs only at $\mathbf{q}=(0,0)$. Also the difference in the QPI pattern between $\tau_{3}$-scattering and $\tau_1$-scattering indicates the importance of 
identifying the nature of impurity scattering when QPI data are interpreted to probe the superconducting pairing symmetry. 

Finally we note that the QPI analysis for the earlier iron pnictide superconductors 
with $s_{\pm}$-wave pairing symmetry~\cite{FWang:2009,YYZhang:2009,
AAkbari:2010} shows the most pronounced scattering at 
$\mathbf{q}=(\pm \pi,0)$ [$(0,\pm \pi)$], which is between the electron (at the $M$ point in the BZ) and hole (at the $\Gamma$ point in the BZ) pockets. For the ``122'' iron selenide superconductors, 
due to the absence of the hole pockets at the BZ center,  the quasiparticle scattering can 
only occur between the electron pockets and the QPI pattern shows pronounced structure
around $\mathbf{q}=(0,0)$ [$(\pm \pi,\pm \pi)$] for either $d_{x^2-y^2}$-wave or isotropic $s$-wave pairing symmetry.

{\em Conclusion.~} We have studied the quasiparticle interference pattern 
due to both $\tau_3$ and $\tau_1$ impurity scattering 
in the recently discovered ``122'' iron selenide superconductors. 
We have shown in the superconducting state that, although both $d_{x^2-y^2}$-wave 
and $s$-wave pairing symmetry lead to nodeless quasiparticle excitations, the QPI pattern 
is strikingly different between the two types of pairing symmetry. In the presence of $\tau_3$ 
impurity scattering, the QPI exhibits strongest scattering with momentum transfer $\mathbf{q}=(\pm \pi, \pm \pi)$ for $d_{x^2-y^2}$-wave pairing symmetry due to the sign change in the superconducting gap across electron pockets at the BZ boundary; while the strongest scattering 
exhibits a ring-like structure centered around both $\mathbf{q}=(0,0)$ and $\mathbf{q}=(\pm \pi, \pm \pi)$ for isotropic $s$-wave pairing symmetry.  In the presence of $\tau_1$ impurity scattering, 
the strongest QPI intensity occurs only at $\mathbf{q}=(0,0)$ for the $d_{x^2-y^2}$-wave pairing 
symmetry, while it occurs at both $\mathbf{q}=(0,0)$ and $(\pm \pi, \pm \pi)$ for isotropic 
$s$-wave pairing symmetry. This analysis shows the sensitivity of the QPI pattern to the nature of impurity scattering. 
The significant contrast in the QPI pattern between the $d_{x^2-y^2}$-wave and the isotropic $s$-wave pairing symmetry in the presence of the same type of impurity scattering
can be very efficient for probing the pairing symmetry in the  ``122'' iron selenide 
superconductors within the Fourier-transform STM technique. 

One of us (J.X.Z) acknowledges useful discussions with A. V. Balatsky. 
This work was supported by U.S. DOE  at
LANL  under Contract No. DE-AC52-06NA25396, the LANL LDRD-DR Program, 
and  the DOE Office of Basic Energy
of Sciences.

\end{document}